\documentclass{article}
\pdfoutput=1
\usepackage{ijcai20}

\usepackage{times}
\usepackage{soul}
\usepackage{url}
\usepackage[hidelinks]{hyperref}
\usepackage[utf8]{inputenc}
\usepackage[small]{caption}
\usepackage{graphicx}
\usepackage{amsmath}
\usepackage{amsthm}
\usepackage{booktabs}
\usepackage{algorithm}
\usepackage{algorithmic}
\newtheorem{Definition}{Definition}
\newtheorem{lemma}{Lemma}
\newtheorem{Rule}{Rule}
\usepackage{amsfonts}
\newcommand{\todo}[1]{{\color{black} #1}}
\usepackage{xcolor}
\definecolor{todo}{rgb}{0.858, 0.188, 0.478}
\usepackage{graphics}
\usepackage{subfigure}
\usepackage{epstopdf}
\usepackage{xcolor,pifont}
\newcommand*\colourcheck[1]{%
  \expandafter\newcommand\csname #1check\endcsname{\textcolor{#1}{\ding{52}}}%
}
\newcommand*\colouracross[1]{%
  \expandafter\newcommand\csname #1across\endcsname{\textcolor{#1}{\ding{56}}}%
}
\colourcheck{blue}
\colourcheck{green}
\colouracross{red}
\usepackage{makecell}
\usepackage{pdfpages}

\title{Unifying Homophily and Heterophily Network Transformation via Motifs}

\author{
Yan Ge$^1$
\and
Jun Ma$^2$\and
Li Zhang$^{1}$\And
Haiping Lu$^1$
\affiliations
$^1$Department of Computer Science, University of Sheffield\\
$^2$Amazon.com, Seattle, WA, USA\\
\emails
\{yge5,  lzhang72,  h.lu\}@sheffield.ac.uk,
junmaa@amazon.com
}

\begin{document}

\maketitle

\begin{abstract}
Higher-order proximity (HOP) is fundamental for most network embedding methods due to its significant effects on the quality of node embedding and performance on downstream network analysis tasks. Most existing HOP definitions are based on either \textit{homophily} to place close and highly interconnected nodes tightly in embedding space or \textit{heterophily} to place distant but structurally similar nodes together after embedding. In real-world networks, both can co-exist, and thus considering only one could limit the prediction performance and interpretability. However, there is no \textit{general and universal} solution that takes both into consideration. In this paper, we propose such a simple yet powerful framework called \underline{\textbf{h}}omophily and \underline{\textbf{h}}eterophliy preserving \underline{\textbf{n}}etwork \underline{\textbf{t}}ransformation (H$^2$NT) to capture HOP that flexibly unifies homophily and heterophily. Specifically, H$^2$NT utilises motif representations to transform a network into a new network with a hybrid assumption via micro-level and macro-level walk paths. H$^2$NT can be used as an enhancer to be integrated with any existing network embedding methods without requiring any changes to latter methods. \todo{Because H$^2$NT can sparsify networks with motif structures}, it can also improve the computational efficiency of existing network embedding methods when integrated. We conduct experiments on node classification, structural role classification and motif prediction to show the superior prediction performance and computational efficiency over state-of-the-art methods. \todo{In particular, DeepWalk-based H$^2$NT achieves 24\% improvement in terms of precision on motif prediction, while reducing 46\% computational time compared to the original DeepWalk.}

\end{abstract}
\section{Introduction}
Networks are ubiquitous in the real world, such as social 
networks~\cite{fortunato2010community}, biological networks \cite{girvan2002community} and traffic networks \cite{asif2016matrix}.
Network embedding (a.k.a. graph embedding) learns low-dimensional latent representations of nodes while preserving the structure and inherent properties of the network. It has been successfully applied in node classification \cite{kipf2017semi}, link prediction \cite{zhang2018arbitrary}, and community detection \cite{wang2017community}.


\begin{figure} [!t] 
\centering 
\vspace{5mm}
\subfigure[Homophily]{\includegraphics[width=27mm,height=25.5mm]{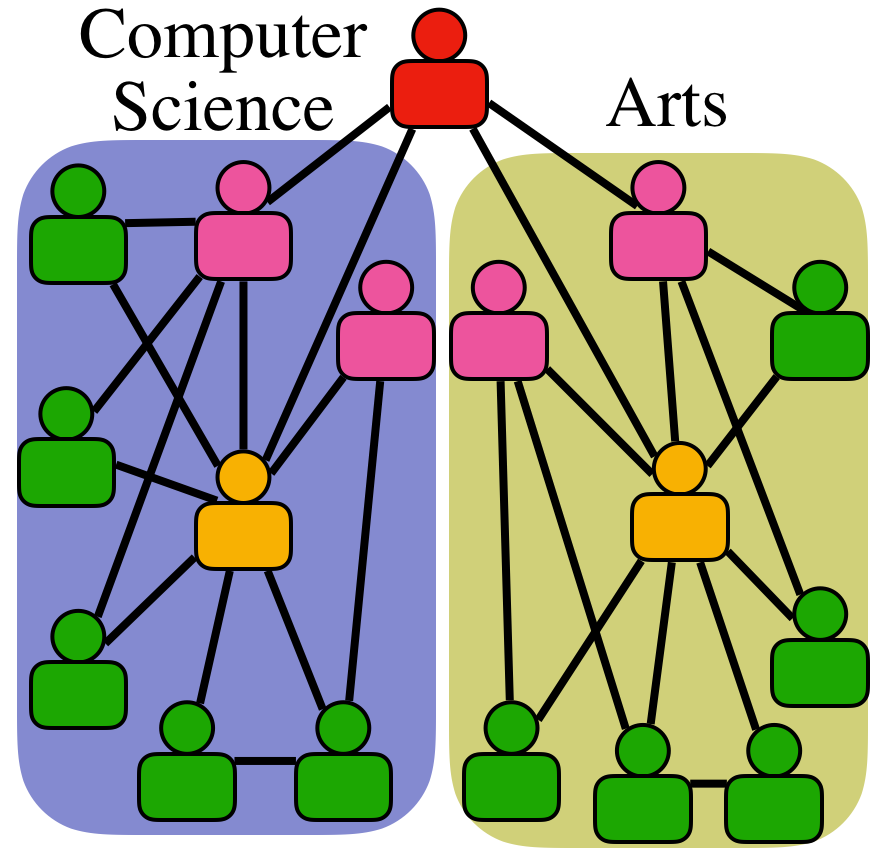}
\label{subfig:homo}}\hspace*{-0.4em}
\subfigure[Heterophily]{\includegraphics[width=27mm,height=25.5mm]{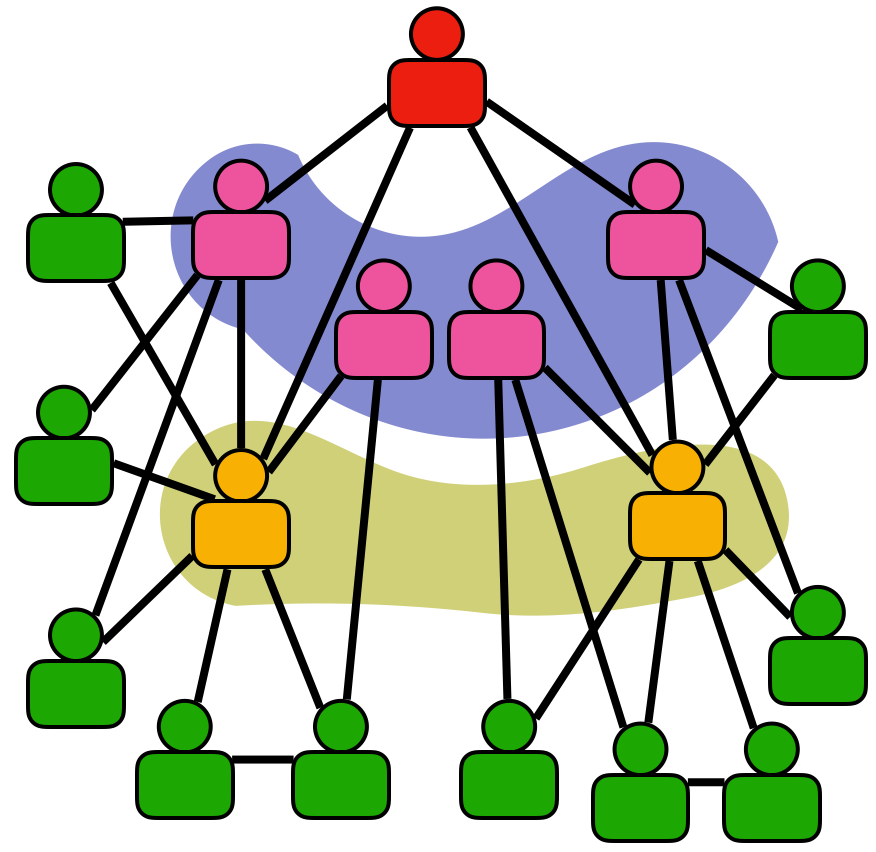}
\label{subfig:heter}}\hspace*{-0.4em}
\subfigure[Hybrid (proposed)]{\includegraphics[width=27mm,height=25.5mm]{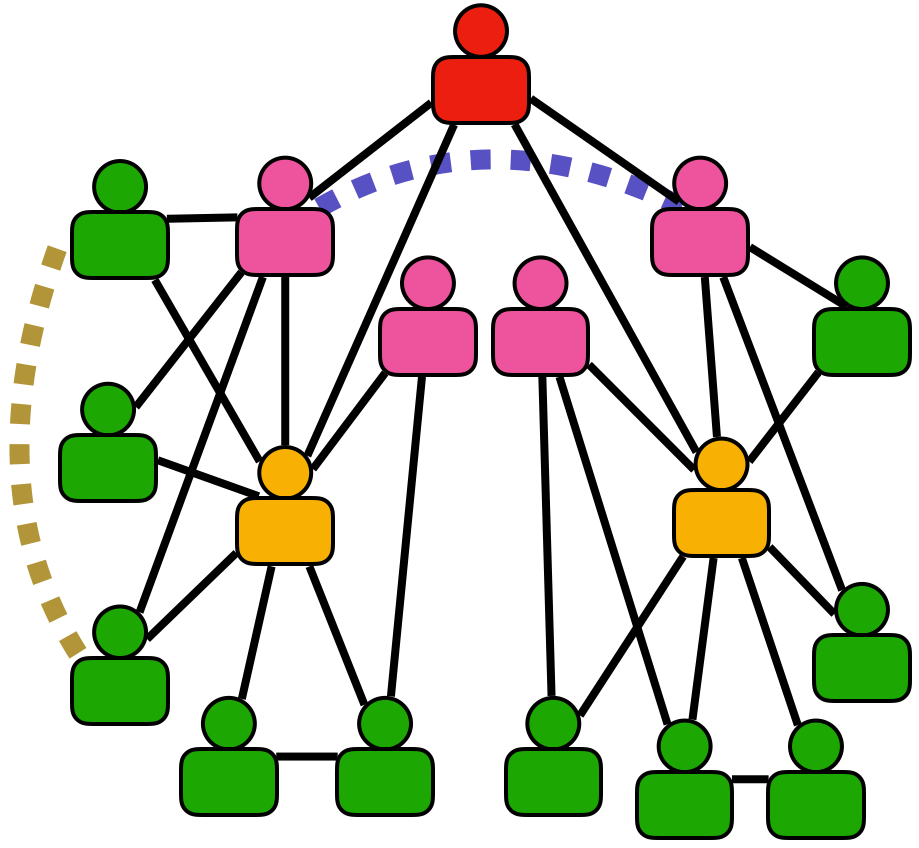}
\label{subfig:hybrid}}
\caption{Three categories of higher-order proximity assumptions for network embedding: homophily, heterophily and hybrid (proposed).
}
\vspace{-7mm}
\label{fig:demo}
\end{figure}

Preserving higher-order proximity (HOP) instead of only considering direct neighbourhood relationship (e.g., adjacency matrix) has been shown to be effective for network embedding since it can
capture rich underlying structures of 
networks \cite{cao2015grarep,tang2015line,zhang2018arbitrary}. Based on the proximity assumption, 
there are three categories of HOP: homophily,
heterophily, and hybrid. In homophily \cite{fortunato2010community}, nodes that are
highly interconnected and in the same community should be placed tightly in embedding space. 
For example, in Fig.~\ref{subfig:homo},  
proximity within the same department should be higher than 
different ones after embedding, which benefits community detection~\cite{wang2016structural} and 
node classification~\cite{perozzi2014deepwalk}. 
For heterophily~\cite{klicpera_diffusion_2019}, nodes that are far away and in different groups, but due to their strong structural similarity, they should be close after embedding.
For example, in Fig. \ref{subfig:heter}, department heads from different academic areas
(yellow nodes) should have stronger relation than their immediate neighbours due to the same job role under heterophily, good for structural role classification  \cite{rossi2019community}. For hybrid assumption, both 
homophily and heterophily can be flexibly preserved, which potentially benefits
(long/short-range) link prediction task (e.g., Fig. \ref{subfig:hybrid}). 

To preserve HOP under the homophily assumption, 
DeepWalk~\cite{perozzi2014deepwalk}
employs random walks to generate node sequences analogous to word sentences, and then the HOP is approximately captured by a
Skip-gram model~\cite{mikolov2013distributed}.
Graph convolutional network (GCN)~\cite{kipf2017semi} aggregates the  features of local neighbours for central node representation so that feature vectors of nodes within the same community are 
more similar than those in different communities.
Arbitrary-order proximity embedding (AROPE) \cite{zhang2018arbitrary} effectively and accurately preserves
arbitrary-order proximity by reweighting the eigen-decompostion.
To preserve HOP under the heterophily assumption,
struc2vec~\cite{ribeiro2017struc2vec} first encodes the node 
structural similarity into a multi-layer graph,
and then DeepWalk is  performed on this multi-layer graph to learn node representations.
 A recent transformation model graph diffusion
 convolution (GDC)~\cite{klicpera_diffusion_2019}
 generates a new network by constructing a diffusion graph obtained by a polynomial function, 
 and then sparsify this diffusion graph by setting a threshold, 
 but still overlooks the 
 heterophily assumption.
 
The higher-order proximities in the aforementioned methods are defined to be
either homophily or heterophily. Such ``one-size-fit-all'' proximity
representation potentially limits the performance and interpretation on many 
network-based tasks. 
 To alleviate this problem, one representative \textit{hybrid} solution is the node2vec method~\cite{grover2016node2vec}, which flexibly adopts both breadth-first and depth-first search strategies to conduct a biased random-walk process. However, it is designed specifically for random-walk-based methods
and cannot take advantage of other more powerful network embedding methods 
proposed recently, such as GCN and AROPE.
Additionally, most existing proximity preserving methods still rely on simple
pairwise relations without considering motifs 
(e.g., triangles, $4$-vertex cliques) 
that can directly capture interactions between more than two nodes \cite{benson2016higher}.
For example, triangular structures, with three reciprocated edges connecting three nodes, play important roles in 
social networks \cite{kossinets2006empirical} \todo{that can be partitioned into
dense triangle  communities with  motif spectral clustering (MotifSC) \cite{benson2016higher}.}
This paper will focus on fundamental triangle
motif structure, though our proposed method can be easily extended to other motifs.

\begin{table}[t!]
	\caption {Comparison with state-of-the-art methods in term of fulfilled (\greencheck)
	and missing (\redacross) properties.}\label{tab:compare}
	\setlength{\tabcolsep}{1.1pt}
	\renewcommand{\arraystretch}{1.1}
 	\resizebox{\columnwidth}{!}{
	\centering
	\large
	\begin{tabular}{lccccccc|c}
		\toprule 
		\textbf{Property} & DeepWalk & LINE&node2vec&AROPE&GCN&struc2vec&MotifSC&\textbf{H$^2$NT}\\ 
		\midrule	
		Homophily &\greencheck	&\greencheck	&\greencheck	&	\greencheck&\greencheck&\redacross&\greencheck&\greencheck\\
		Heterophily &\redacross	&\redacross	&\greencheck	&\redacross	&\redacross&\greencheck&\redacross&\greencheck\\
		Motif &\redacross	&\redacross	&\redacross	&\redacross	&\redacross&\redacross&\greencheck&\greencheck\\
		\bottomrule
	\end{tabular}
 	}
 	\vspace{-4mm}
\end{table}
To design a general framework with a hybrid HOP assumption,  we propose a
    \underline{\textbf{h}}omophily and \underline{\textbf{h}}eterophliy preserving \underline{\textbf{n}}etwork \underline{\textbf{t}}ransformation (H$^2$NT) with motif representations. Our H$^2$NT defines a new HOP by micro-level
and macro-level walk paths as two complementary components
to represent homophily and heterophily.
    The \textit{micro-level} walk paths embody the homophily assumption, aiming to collect the similarity of close neighbours 
according to their homophily levels generated by motif information. 
 The \textit{macro-level} walk paths embody the heterophily assumption, aiming to encourage walk paths to explore global information according to structural similarity.

 As a general framework, H$^2$NT is not limited to one specific algorithm but can be integrated to any network embedding algorithm as a preprocessing step, without requiring changing their cores. 
Furthermore, the two walk path strategies can only  rely  on  local  motif  structures  and  sparsify  networks and subsequently improve the computational efficiency when we integrate H$^2$NT with existing network embedding algorithms. 
Table \ref{tab:compare} compares three desirable properties to show the uniqueness of H$^2$NT compared with several state-of-the-art (SOTA) methods. 
\todo{To summarise, the contributions of our paper are as follows:
\begin{enumerate}
\item We propose micro-level and macro-level walk paths to preserve homophily and heterophily in HOP by theoretically studying why most HOP preserving embedding methods only hold a homophily assumption.
\item We propose a simple and novel framework to unify homophily and heterophily representations according to micro-level and macro-level walk paths, and three instantiations.
\item  We conduct experiments on three tasks, node classification, structural role classification, and motif prediction (a generalised link prediction problem) to show the superior performance over SOTA methods.
\end{enumerate}
}

\section{Preliminary}

\subsection{Notations.}
We denote scalars by lowercase letters, e.g., $d$, 
vectors by lowercase boldface letters, e.g., $\mathbf{d}$, matrices by  uppercase  boldface, e.g., $\mathbf{D}$. 
Let $G = (V, E)$ be an undirected unweighted graph (network) with $V$ = \{v$_1$, v$_2$, \dots, v$_n$\} being the set of $n$ nodes, i.e., $n$ = $|V|$, and $E$ = \{e$_1$, e$_2$, \dots, e$_n$\} being the set of edges connecting two nodes. 


\subsection{Higher-Order Proximity.}
Network embedding aims to learn latent, low-dimensional representation of nodes while preserving network 
topology~\cite{zhang2018network}. Prior works have demonstrated that,  the higher-order proximities between nodes are of tremendous importance in capturing the underlying structure of the network~\cite{cao2015grarep,ou2016asymmetric,yang2017fast,feng2018representation}. 
 The adjacency matrix $\mathbf{A}$ can be treated as the first-order proximity, which captures the
pairwise proximity between nodes. However, the first-order proximity is very sparse
and insufficient to fully model the relationships between nodes in most cases. In order to
characterise the connections between nodes better, HOP is widely studied.
Given $\mathbf{A}$, an HOP can be defined as a polynomial function of $\mathbf{A}$ \cite{zhang2018arbitrary}:
$
\mathbf{S}=w_{1} \mathbf{A}+w_{2} \mathbf{A}^{2}+\ldots+w_{l} \mathbf{A}^{l},
$
where $l$ is the order, and $w_1 , ..., w_l$ are the weights for each term. Matrix $\mathbf{A}^l$ denotes the $l$th-order proximity matrix, with multiplication of $l$ matrices $\mathbf{A}$. 
The $l$th-order proximity value
between nodes $v_i$ and $v_j$ is denoted as 
$a_{ij}^{(l)}$. 

\section{Methodology}

\subsection{Theoretical Framework and Motivations}
\todo{In this section, we theoretically reveal 
why most network embedding algorithms only hold a homophily assumption in HOP. This motivates us to propose  micro-level  and  macro-level  walk  paths strategies to  represent  homophily  and  heterophily in HOP.}
\label{sec:theory}

We first introduce a fully-connected planted partition model (PPM) as follows \cite{condon2001algorithms}:
\begin{Definition}
(\textbf{Fully-Connected PPM}) \ Let $G_f \sim G_f(mr, r, p, q)$ be a graph sampled from the planted partition model on $mr$
vertices, with $r$ clusters $C = 
\{C_1, \cdots, C_i, \cdots, C_r\}$ each with exactly $m$
vertices. The edge set is then generated as follows:
two vertices $\{v_i,v_j\}\in C_i$ are connected with weight $p$ otherwise with
weight $q<p$ to ensure well-connected clusters.
\end{Definition}

We prove the following lemma to show a relationship between homophily and HOP.
\begin{lemma}\label{lemma:relation}
Let $G_f \sim G_f(mr, r, p, q)$ be a fully-connected PPM  with $r$ clusters
$C = \{C_1, \cdots, C_i, \cdots, C_r\}$, nodes \{ $v_i,v_j\}\in C_s$ and 
$v_k \in C_t$, the value of $l^{th}$-order proximity between $v_i$ and $v_j$  is $a_{ij}^{(l)}$, 
then
$
a_{ik}^{(l)} < a_{ij}^{(l)} \label{equ:ADJ}.
$
\end{lemma}
\todo{Our proof as shown in Supplementary Material is based on the following rule.}
\begin{Rule}
\textbf{(Chapman-Kolmogorov equations)} \cite{tijms2012understanding},\vspace{-7mm}
\end{Rule}

\begin{equation}\label{rule:CK}
a_{ik}^{(l)} = \sum_{z=1}^{kn}a_{iz}^{(l-1)} \cdot a_{zk},\ \ 
a_{ij}^{(l)} = \sum_{z=1}^{kn}a_{iz}^{(l-1)} \cdot a_{zj}.
\end{equation}

\textbf{Observations.} 
We have two main observations:
1)  From Lemma \ref{lemma:relation},
we see that most existing HOPs hold the assumption of homophily, so that
in the embedding space, distance of two nodes residing in different communities being inherently larger than that of those in the same community.
2)  Eqs (\ref{rule:CK}) reveal that HOP between any pair of nodes $v_i$ and $v_j$
essentially represents the total similarity of a sequence of
nodes traversed by all possible walk paths from $v_i$ to $v_j$,
\todo{and all walk paths share the same contributions to represent $a_{ij}^{(l)}$ regardless of various downstream  network tasks. However, we argue that every
walk path should have task-relevant contributions to $a_{ij}^{(l)}$.} 
This inspires the following question: \textit{are there some specific walk paths that characterise the \todo{task-relevant} HOP w.r.t. homophily or heterophily?}  Subsequently, 
to explicitly reveal the characteristics of HOP, we categorise all possible walk paths into micro-level and macro-level walk paths as two complementary components to represent HOP. Our proposed definitions are below.

\begin{Definition}(\textbf{Micro-level walk path})
A micro-level walk path connecting $v_i$ and $v_j$ is a sequence of 
vertices $V'= ($$v_i$, $v_k$,$\cdots$, $v_j$) traversed a sequence of 
edges 
$E'= ($$e_1$, $e_1$,$\cdots$, $e_n$) that ensures $V'\subseteq C_i$ and 
$E'\subseteq C_i$.
\end{Definition}

\begin{Definition}
(\textbf{Macro-level walk path}) A macro-level walk path connecting $v_i$ and $v_j$ is a sequence of vertices
$V'= ($$v_i$, $v_k$,$\cdots$, $v_j$) traversed a sequence of edges 
$E'= ($$e_1$, $e_1$,$\cdots$, $e_n$) and $E' \cap C_j \neq \phi$ and $E' \cap C_i \neq \phi$ and $i \neq j$.
\end{Definition}

\begin{figure}[t]
\includegraphics[width=180pt]{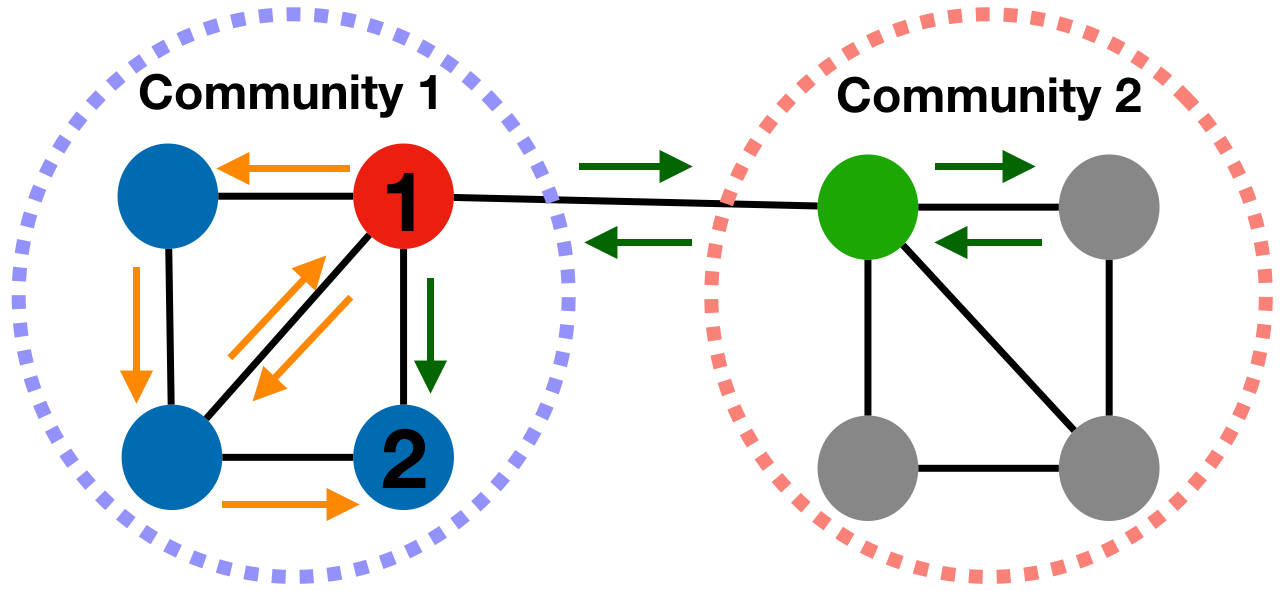}
\centering
\caption{Illustration: 1) Micro-level (orange arrows \todo{within the community 1}) and macro-level (dark green arrows \todo{across communities 1 and 2}) walk paths represent the 5th-order proximity between nodes 1 and 2; 2) The green node contributes more centrality to the red node than any blue node.
}
\label{fig:path}
\vspace{-3mm}
\end{figure}
For example, in Fig \ref{fig:path}, we observe that 
orange (micro-level) path and dark-green (macro-level) path 
play very different roles to induce a five-step connectivity pattern from node 1 to node 2, although both can contribute to
5th-order proximity between node 1 and 2. Under the homophily assumption, the orange path
has more expressive power than the dark-green path since it is likely
to leverage tightly close neighbourhood similarity within a community,
\todo{which may benefit community detection and node classification tasks.}
In contrast, under the heterophily assumption,
the dark-green path
has more appropriate expressive power than the orange path to heterophily since it tends to use weak-connectivity and distant neighbourhood similarity across communities, \todo{good for structural role classification.  Moreover, only homophily or heterophily cannot be suitable for all network-based tasks. }

\todo{In general, micro/macro-level walk paths allow us to capture the structure of the traversed region and provide an attention mechanism to guide the walk. This allows us to focus on task-relevant parts of the graph while eliminating the noise in the rest of the graph which results in the network embedding that provides better predictive performance.} Note that different from the breadth-first and depth-first 
approximately search in node2vec, 
fixed length of micro-level and macro-level walk paths will be exhaustedly and accurately
considered to represent
HOP between two nodes. While they share a general idea that representation of a node is determined by its neighbourhoods that need to be flexibly defined. 

Building on the above discussions, in the following
we propose H$^2$NT that defines new HOP to flexibly preserve both homophily and heterophily by characterising walk paths with micro-level and macro-level walk paths. It is a generic model that can be used as a preprocessing step to provide input to 
any network embedding methods.

\subsection{Homophily Proximity Representation}
To represent homophily proximity, adjacency matrix $\mathbf{A}$ or random walk matrix $\mathbf{P}$ is widely used but both are noisy and sparse. Another choice is to perform community detection so that homophily character of each edge can be clearly shown, such as by spectral clustering~\cite{ng2002spectral}). However, this will lead to a binary-valued output
edge (i.e., within or across communities) and limits our ability in exploring unseen patterns of a network. 

To address the above challenges, we adopt a scalable and \textit{smooth} community detection method with 
motif representations \cite{benson2016higher} to represent  
homophily proximity as $\mathbf{A}_M$, 
\begin{equation} \label{eq:motif-count}
\mathbf{A}_M(i, j)=\sum_{v_{i}, v_{j} \in \mathcal{V}}
\underline{1}\left(v_{i}, v_{j} 
\operatorname{occurin} M\right)
\end{equation}
where $i\neq j$, $v_i$ and $v_j$ belong to motif $M$ , and \underline{1}($s$) is the truth-value
indicator function, i.e., \underline{1}($s$) = 1 if the statement s is true and 0
otherwise. Note that the weight is added to $\mathbf{A}_M(i, j)$ only if node
$v_i$ and $v_j$ occur in the given motif $M$. In this paper, we only focus on
undirected triangle motif, but it can be easily generated to any other type of motifs.
The intuition of homophily proximity representation 
(Eq.~(\ref{eq:motif-count})) is that motif representation smooths out the
neighbourhood over the graph,  acting as denoise filter due to removed edges that
do not participate in any motif.
More importantly, the value in homophily proximity representation indicates the level of homophily as quantified in 
Lemma~\ref{lemma:motif_homo} 
extended from~\cite{tsourakakis2017scalable} \todo{and our proof as shown in Supplementary Material:}

\begin{figure}[t]
\includegraphics[width=0.5\textwidth]{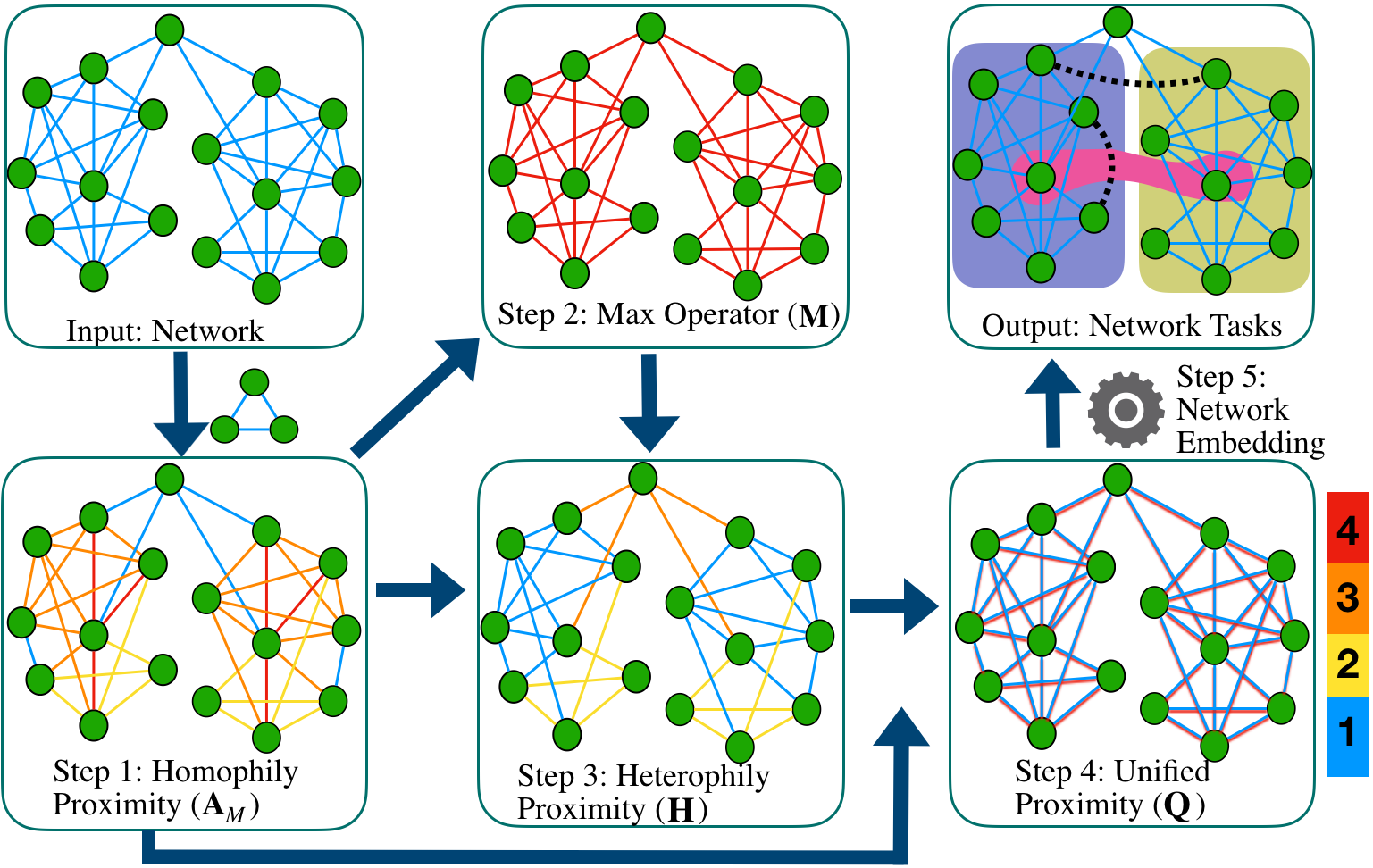}
\centering
\caption{The H$^2$NT framework. The colour bar indicates the weight scales
of edges. (See text for details.)}
\label{fig:frame}
\vspace{-3mm}
\end{figure}

\begin{lemma} \label{lemma:motif_homo}
Let $G \sim G(mk, k, p, q)$ be an unweitghted graph sampled by a
generalized PPM, $\{v_i,v_j\}\in C_i$ and 
$\{v_k\} \in C_j$, $\{v_d\} \in C_b$, $j\neq b$ and $e(v_i,v_j) \in E, e(v_d,v_k)\in E$. 
We use $\mathbb{E}[w(v_i, v_j)]$ to indicate the expected number of triangles  containing the edge $\{v_i,v_j\}$, then 
$
\mathbb{E}[w(v_d, v_k)] < \mathbb{E}[w(v_i, v_j)].
$
\end{lemma}





 \todo{Lemma \ref{lemma:motif_homo} shows the effectiveness of
 homophily proximity by the triangular motif.} Specifically, 
 an edge within a community will contain more triangles than that of edge cross communities under the planted partition model
 and then homophily proximity is naturally revealed in this constructed motif graph. 
\todo{ The number of triangles containing
 an edge $e(v_i, v_j)$ indicates the level of contributions to represent homophily proximity between $v_i$ and $v_j$.} 
Leveraging this idea further, it will help us characterise whether a walk path 
is likely to explore local, close neighbourhoods or global far-away nodes. 



\subsection{Heterophily Proximity Representation}
\todo{Most of existing network embedding algorithms (e.g., DeepWalk, AROPE)
hold a homophily assumption in HOP but overlook heterophily.}
To discover heterophily proximity, we encourage macro-level walk paths
to contribute more to proximity since it reflects the similarity of neighbourhoods
in global position. 
However, estimation of  macro-level walk paths
seems non-straightforward to overcome \todo{because heterophily information is not explicitly 
provided in most common networks. A signed network 
$G^{\pm}=(G^{+}, G^{-})$ has both positive and negative edge weights, where positive relations $G^{+}$
encode friendship, and negative relations $G^{-}$ encode enmity 
interactions~\cite{leskovec2010signed}. Essentially, a signed network
is represented by two networks with largely different structural properties.
} 

\todo{Inspired by signed networks, we 
propose motif-aware signed networks $G_M^{\pm}=(G_M^{+}, G_M^{-})$. where $G_M^{+}$ and $G_M^{-}$  are
homophily and heterophily networks respectively. An intuition of designing $G_M^{\pm}$ is  based on a random walk theory  that 
homophily/heterophily edges increase the probability of staying in/escaping from a community.
We represent homophily as $\mathbf{A}_M$ shown in the last section.
We represent 
heterophily as $-\mathbf{A}_M$ by turning all values in $\mathbf{A}_M$ into their opposite numbers.
We interpret that the larger $-\mathbf{A}_M (i, j)$, the higher chance to achieve macro-level walk paths that explore global structures. However, most network embedding methods do not have capacity to handle negative
weights. To develop a general and universal network transformer model, we use the following simple but efficient approach to transit negatives into positives while preserving heterophily:
}
\begin{equation} \label{eq:hetero}
    \todo{\mathbf{H} = -\mathbf{A}_M + \mathbf{M},}
\end{equation}
where
\begin{equation*}\label{adj-tensor}
\mathbf{M} (i, j)=\left\{
\begin{aligned}
f_{max}(\mathbf{A}_M) &&  \textrm{$v_i, v_j$  are contained in a triangle},\\
0 && \textrm{otherwise,}
\end{aligned}
\right.\end{equation*}
where $f_{max}(\mathbf{A}_M)$ indicates a maximum value in $\mathbf{A}_M$.
Heterophily proximity ($\mathbf{H}$) can be achieved by Eq. (\ref{eq:hetero}).

Building upon~\cite{alvarez2015eigencentrality} 
and~\cite{zhang2018arbitrary},
heterophily proximity ($\mathbf{H}$) is a proper representation of node 
centrality as an application of structural similarity. Heuristically, 
contribution centrality is a centrality measure that 
has more influence on centrality 
from one given node to central node with greater contribution centrality. 
The influence on centrality can be quantified with closeness centrality~\cite{estrada2005subgraph} that is the number of times a node acts as a bridge along the shortest path between two other nodes. 
Heterophily proximity permits us to quantify the contribution centrality.
For example, regarding Fig. \ref{fig:path} as a heterophily proximity network, 
considering centrality of red node, 
centrality contribution from green to red is larger than that from any immediate blue nodes, which shows intuitively our heuristic analysis.
 The reason is that the red one can access the grey ones only through the green, and the blue
 nodes are redundant for the red one, because it can access directly each blue node without any
intermediary. Therefore, our heterophily proximity can efficiently encode contribution centrality.




\subsection{Unification and Instantiations}
\label{secunine}
We linearly unify the homophily and heterophily proximity as follows: 
\begin{equation}
\mathbf{Q} = \mathbf{A}_M + \lambda\mathbf{H},
\end{equation}
However, micro-level walk paths are still able to represent proximity with increasing
of orders due to the inherit graph structure. To prevent it, we need to enlarge the difference of entries in $\mathbf{H}$. Considering the scalability issue in the network embedding field, we focus on 
a linear combination ($\lambda$), but similar ideas (e.g., 
non-linear operators) can
be straight-forwardly generalised.
The first-order proximity $\mathbf{Q}$ flexibly shows homophily 
and heterophily characteristics by hyperparameter $\lambda$ that 
controls over the importance of heterophily. Fig~\ref{fig:frame} illustrates the proposed H$^2$NT framework. 

Our framework H$^2$NT can be integrated with any network embedding methods as a preprocessing step to provide input to them. Here we select three representative algorithms,  AROPE on matrix factorisation, DeepWalk on random walks and 
GCN on convolutional neural networks. 


\begin{itemize}
  \item \textbf{H$^2$NT-AROPE} (H$^2$NT-A). 
To preserve HOP, we use a linear combination of power of biased matrix $\mathbf{Q}$ as follows,
      $\mathbf{P}_M=w_{1} \mathbf{Q}+w_{2} \mathbf{Q}^{2}+\ldots+w_{l} \mathbf{Q}^{l}.$
  When $\lambda=0$, we interpret  $\mathbf{Q}^r (i, j)$ ($1\leq r \leq l$) as the total number of motifs traversed by all possible $l$-length
  walk paths connecting nodes $v_i$ and $v_j$. Increasing $\lambda$ results in more similarity from global neighbours to represent
  the HOP between $v_i$ and $v_j$.
  Moreover, benefiting from AROPE, H$^2$NT-A can 
  explore arbitrary walk length $l$ between two nodes without increase computational complexity. 
  \item \textbf{H$^2$NT-DeepWalk} (H$^2$NT-D). 
  The objective function of DeepWalk can be written as:
  \begin{equation*}
      \max_{\Phi}  \ \log \text{Pr}\left(\left\{v_{i-w},\dots,v_{i-1}, v_{i+1},\dots,v_{i+w}\right\} | \Phi(v_i)\right), \label{eq:DW-obj}
  \end{equation*}
  where $w$ is the window size, $\Phi(v_i)$ is the representation of $v_i$.
  Under the proposed H$^2$NT framework, instead of uniformly random sampling neighbours of $v_i$,
  we adopt a biased sampling strategy. Different from the biased sampling strategy in node2vec that   considers a second-order random walks by tuning two out-in parameters, H$^2$NT-D only needs to tune one unifying parameter $\lambda$. A small $\lambda$ makes it more likely to sample local, close neighbours. By contrast, 
  a large $\lambda$ makes it more likely to sample global, far-away neighbours.
  
  \item \textbf{H$^2$NT-GCN} (H$^2$NT-G). The layer-wise propagation rule in GCN can be written as,
  \begin{equation}\label{eq:GCN}
      \mathbf{H}^{(l+1)}=\sigma\left(\tilde{\mathbf{D}}^{-\frac{1}{2}} \tilde{\mathbf{Q}} \tilde{\mathbf{D}}^{-\frac{1}{2}} \mathbf{H}^{(l)} \mathbf{W}^{(l)}\right),
  \end{equation}
  where $\tilde{\mathbf{Q}}=\mathbf{Q}+\mathbf{I}_N$, $\mathbf{I}_N$ is an identity matrix, $\tilde{\mathbf{D}}_{i i}=\sum_{j} \tilde{\mathbf{Q}}_{ij}$,  $\mathbf{W}^{(l)}$ is a layer-specific trainable weight matrix, $\mathbf{H}^{(l)}$ is an activation in the $l^{th}$ layer; $\mathbf{H}^{(0)}$ is the given feature matrix, and $\sigma(\cdot)$ is an activation function. From Eq. (\ref{eq:GCN}), 
  H$^2$NT-G can discriminate 
  the neighbourhoods. Specifically, 
  it will give more attention to neighbourhoods that has homophily
  assumption if $\lambda$ is small, and otherwise, it will give more attention to neighbourhoods that has heterophily
  assumption.  
\end{itemize}
\textbf{Complexity Analysis.}
The time complexity of homophily representation can be as large as $O(n^3)$ for a complete graph, where $n$ is the number of nodes in the network.
Let $t$ is the number of triangle of a input network $G$. While most
real networks are far from complete so the actual complexity is much lower than $t < O(n^3)$. According to empirical study in \cite{benson2015tensor}, the value of $t$ in real-world 
networks is linear with $|E|$.
For heterophily and unification step, each has the same complexity, which is the number of
non-zero entries $j \leq |E|$ in $\mathbf{A}_M$. Only $j = |E|$ when the network 
is complete.
Thus the total complexity of H$^2$NT is $O(n^3)$+$O(j)$
after ignoring lower order terms.

\section{Experiments}
\textbf{Datasets.} We conduct extensive experiments on the following seven real networks covering social networks and traffic networks: 1) Amherst, Hamilton, Mich, Rochester \cite{traud2012social}
are the Facebook social networks at different universities in US. 
We use \textit{class year} as the \textit{node labels} in \textit{node classification}.  
2) Brazil, Europe, USA are air-traffic networks from \cite{ribeiro2017struc2vec}. Nodes indicate airports and edges correspond to commercial airlines. 
We use the level of airport activity (e.g., passenger traffic) as the node labels
in structural role classification. Statistics of networks are shown in Table \ref{statistics}.


\begin{table}[t]
	\caption {Statistics of networks with isolated nodes removed. \#Test Triangles indicates the number of removed triangles as testing set for motif prediction.}\label{statistics}
	\renewcommand{\arraystretch}{1}
	\resizebox{\columnwidth}{!}{
	\centering
	\begin{tabular}{lrrrr|cc}
		\toprule 
		\textbf{Network} & $|V|$ & $|E|$&Edge Density&Labels&\#Test Triangles\\ 
		\midrule
		 Amherst&2,021&81,492& 40.3&15&10K\\
		 Hamilton&2,116&87,486& 41.3&15&10K\\
		Mich &2,924&54,903& 18.7&13&10K\\
		Rochester &4,140&14,5309& 35.1&19&10K\\ \hline
		Brazil&131&1,038& 7.9&4&200\\
		Europe&399&5,995& 15.0&4&300\\
		USA &1,190&13,599& 11.4&4&500\\
		\bottomrule
	\end{tabular}
 	}
 	\vspace{-3.5mm}
\end{table}

\textbf{Baselines.} We extensively compare the proposed H$^2$NT with the following eight 
state-of-the-art methods covering network embedding methods and a SOTA network transformer method:
1) Deepwalk\footnote{https://github.com/phanein/deepwalk}: vary window size\{1, 2, 3, 4, 5, 6\}
and use default settings for other hyperparameters.
2) LINE\footnote{https://github.com/snowkylin/line}:  study two versions of LINE that preserves the first-order proximity (LINE-1st) and  second-order proximity (LINE-2nd). We use the default settings for other hyperparameters.
3) node2vec\footnote{https://github.com/aditya-grover/node2vec}: vary the bias hyperparameters inward, outward from \{0.25, 0.5, 1, 2, 4\} and use the default settings for other hyperparameters.
4) AROPE\footnote{https://github.com/ZW-ZHANG/AROPE}:  tune the number of preserved higher-order proximity \{1, 2, 3, 4, 5, 6\}  and $w_i$ = 0.1$^i$.
5) struc2vec\footnote{https://github.com/leoribeiro/struc2vec}:  study all four different optimisation strategies of struc2vec.
6) Graph Neural Network\footnote{https://github.com/tkipf/pygcn}:
 use the default settings to conduct experiments.
7) MotifSC:  use the default settings to conduct experiments.
8) GDC \footnote{https://github.com/klicperajo/gdc}:  
study two variants of GDC, 
and combine GDC with AROPE (GDC-A), DeepWalk (GDC-D) and GCN (GDC-G) with recommended transport probability \{0.05, 0.15, 0.3\}, exponential in heat kernel \{1, 5, 10\}, and sparsity threshold \{10$^{-5}$, 10$^{-4}$, 
10$^{-3}$, 10$^{-2}$\}.
For our method, we study three variations: H$^2$NT-A, H$^2$NT-D and H$^2$NT-G. 
H$^2$NT-A and H$^2$NT-D take the number of proximity as \{1, 2, 3, 4, 5, 6\} and $\lambda$ = \{0.1, 0.3, 0.5, 0.7, 1.3, 1.5, 1.7\}. H$^2$NT-G uses the same  $\lambda$ with other variant but only uses two layers. 

The dimension of embedding vector is 128 for all social networks and 16 for all traffic networks considering the number of nodes in graphs.
The best performance results for all methods will be reported, with termination of 
the computation if no complete result is returned within twelve hours. 
We use the open-source Python library GEM\footnote{https://github.com/palash1992/GEM} \cite{goyal2018graph} to study all methods
under the same software framework. All experiments
were performed on a Linux machine with 2.4GHz Intel Core and 16G memory.
 We will release the code for H$^2$NT.

\begin{table}[t!]
	\caption {Motif prediction (i.e. generalised
link prediction) results reported in precision @$N_p$. The best results are in \textbf{bold} and the second best ones are \underline{underlined}. 
We set $N_p$ to 500 for all traffic networks and 10k for all Facebook social networks.}\label{tab:motif-pred}
	\setlength{\tabcolsep}{1.4pt}
	\renewcommand{\arraystretch}{1}
 	\resizebox{\columnwidth}{!}{
	\centering
	\large
	\begin{tabular}{lccccccc}
		\toprule 
		Methods &Amherst&Hamilton&Mich&Rochester& Brazil &USA&Europe\\ 
		\midrule	
		MotifSC &0.658	&0.654	&0.702	&0.873	&0.096	&0.187	&0.331\\
		AROPE &\underline{0.894}	&\underline{0.898}	&\underline{0.928}	&0.955	&\underline{0.548}	&\underline{0.985}	&\underline{0.742}\\
		DeepWalk &0.639	&0.658	&0.789	&0.864	&0.050	&0.060	&0.035\\
		LINE-1st &0.082	&0.084	&0.085	&0.088	&0.066	&0.088	&0.082\\
		LINE-2nd &0.310	&0.310	&0.307	&0.319	&0.075	&0.062	&0.060\\
		node2vec &0.094	&0.085	&0.087	&0.097	&0.125	&0.100	&0.295\\
		struc2vec &0.167	&0.181	&0.225	&0.152	&0.387	&0.628	&0.126\\
		GDC-A &0.679	&0.734	&0.665	&0.774	&0.515	&0.881	&0.635\\
		GDC-D &0.821	&0.811	&-	&-	&0.121	&0.397	&0.282\\
		\hline \hline
		H$^2$NT-A &\textbf{0.928}	&\textbf{0.927}	&0.927	&\textbf{0.978}	&\textbf{0.578}	&\textbf{0.986}	&\textbf{0.747}\\
		H$^2$NT-D &0.864	&0.857	&\textbf{0.938}	&\underline{0.972}	&0.098	&0.777	&0.324\\
		\bottomrule
	\end{tabular}
 	}
 	\vspace{-3.5mm}
\end{table}

\begin{table*}[t!]
	\caption {Node classification results (accuracy) on three datasets. The best results are in \textbf{bold} and the second best ones are \underline{underlined}. \todo{The results of LINE-1st and LINE-2nd have much lower accuracy so they are not shown. }  }\label{tab:classification}
	\renewcommand{\arraystretch}{0.9}
 	\resizebox{\textwidth}{!}{
	\centering
	\begin{tabular}{lccccc|ccccc|cccccc|}
		\toprule 
		\multicolumn{1}{c}{}&\multicolumn{5}{c|}{Hamilton}&\multicolumn{5}{|c|}{Rochester}&\multicolumn{5}{|c}{Mich}\\ \hline
		\textbf{\%Lables} & 2\% & 4\%&6\%&8\%&10\%& 2\% & 4\%&6\%&8\%&10\%& 2\% & 4\%&6\%&8\%&10\%\\ 
		\midrule	
		MotifSC &0.209	&0.218	&0.255	&0.249	&0.327	&0.210	&0.221	&0.244	&0.279	&0.303	&0.208	&0.233	&0.231	&0.239	&0.241\\
		AROPE&\underline{0.770}	&\underline{0.840}	&\underline{0.862}	&\underline{0.874}	&\underline{0.876}	&\underline{0.717}	&\underline{0.770}	&\underline{0.790}	&\underline{0.799}	&\textbf{0.811}	&0.465	&\underline{0.506}	&0.524	&\underline{0.537}	&0.542\\
		DeepWalk&0.721	&0.804	&0.842	&0.861	&0.864	&0.711	&0.768	&0.787	&0.798	&0.807	&0.464	&\underline{0.506}	&0.523	&0.536	&\underline{0.547}\\
		node2vec&0.277	&0.317	&0.331	&0.348	&0.354	&0.252	&0.292	&0.319	&0.334	&0.340	&0.226	&0.247	&0.258	&0.257	&0.257\\ 
		GCN&0.649	&0.679	&0.704	&0.742	&0.747	&0.605	&0.660	&0.650	&0.680	&0.675	&0.422	&0.494	&\textbf{0.531}	&\textbf{0.549}	&\textbf{0.550}\\
		struc2vec&0.218	&0.238	&0.251	&0.260	&0.271	&0.201	&0.207	&0.211	&0.214	&0.215	&0.196	&0.208	&0.220	&0.224	&0.225\\ 
		GDC-D	&0.758	&0.829	&0.848	&0.859	&0.862	&0.690	&0.738	&0.756	&0.765	&0.773	&\underline{0.475} &0.505	&0.513	&0.524	&0.532\\
		GDC-A&0.224	&0.284	&0.317	&0.313	&0.423	&0.212	&0.238	&0.274	&0.332	&0.389	&0.211	&0.237	&0.247	&0.242	&0.251\\
		GDC-G&0.481	&0.592	&0.660	&0.730	&0.719	&0.187	&0.421	&0.269	&0.431	&0.413	& 0.260	&0.329	&0.380	&0.477	&0.409\\\hline\hline
		H$^2$NT-D &\textbf{0.789}	&\textbf{0.843}	&\textbf{0.866}	&\textbf{0.877}	&\textbf{0.879}	&\textbf{0.745}	&\textbf{0.780}	&\textbf{0.796}	&\textbf{0.802}	&\textbf{0.811}	&\textbf{0.485}	&\textbf{0.507}	&\underline{0.525}	&0.535	&0.541\\
		H$^2$NT-A &0.729	&0.795	&0.815	&0.824	&0.829	&0.680	&0.723	&0.744	&0.744	&0.754	&0.438	&0.450	&0.460	&0.470	&0.465\\
		H$^2$NT-G  &0.642	&0.688	&0.730	&0.738	&0.745	&0.605	&0.646	&0.662	&0.674	&0.669	&0.412	&0.484	&0.511	&0.529	&0.533\\
		\bottomrule
	\end{tabular}
 	}
\end{table*}

\textbf{Evaluation Metrics.} 
For motif prediction, we use precision@$N_{p}$ to evaluate the performance \cite{wang2016structural,zhang2018arbitrary}. It is defined as:
$\text { precision @}  N_{p}= \frac{1}{N_{p}} \sum_{i=1}^{N_{p}} \delta_{i},$
where $\delta_{i}$ =1 means the $i$-th reconstructed motif is correct (i.e., the reconstructed motif exists in the network), $\delta_{i}$ =0 otherwise
and $N_{p}$ is the number of evaluated motifs.
For node and structural role classification, we use accuracy, i.e. the percentage of nodes whose labels are correctly classified, to evaluate the performance \cite{zhang2018network}:
$\text {Accuracy} (y,\hat{y})=\frac{1}{n} \sum_{i=1}^{n} \mathbb{I}\left(\hat{y}_{i}=y_{i}\right),$
where $\hat{y}$ and $y$ are predicted label and true label respectively, $\mathbb{I}$ is an indicator operator (1 if two labels are equal otherwise 0).

\begin{table*}[t!]
	\caption {Structural role classification results on Brazil, Europe and USA with 90\% training data. The results of LINE-1st and LINE-2nd have much lower accuracy so they are not shown. The best results are in \textbf{bold} and the second best ones are \underline{underlined}. }\label{tab:role-class}
	\setlength{\tabcolsep}{1pt}
	\centering
	\renewcommand{\arraystretch}{0.8}
 	\resizebox{0.85\textwidth}{!}{
	\large
	\begin{tabular}{lccccccccc|cccccc}
		\toprule 
		Datasets &MotifSC&AROPE&DeepWalk&node2vec&GCN&struc2vec&GDC-D&GDC-A&GDC-G& H$^2$NT-D&H$^2$NT-A&H$^2$NT-G\\ 
		\midrule	
		Brazil &0.564	&\underline{0.686}	&0.429	&0.450	&0.379	&\textbf{0.736}&0.607	&0.436	&0.428	&0.514	&0.664	&0.500\\
		Europe &0.365	&0.535	&0.365	&0.422	&0.362	&\underline{0.568}	&0.530	&0.452	&0.350&0.430	&\textbf{0.577}	&0.450	\\
		USA &0.379	&0.589	&0.493	&0.479	&0.549	&\underline{0.608}&0.588	&0.519	&0.403	&\textbf{0.629}	&0.600	&0.565	\\
		\bottomrule
	\end{tabular}
 	}
 	\vspace{-4.5mm}
\end{table*}

\textbf{Motif Prediction.}
Besides links, motifs are small subgraphs fundamental in networks. Thus, prediction of motif structures is important in real applications. Therefore, we design the motif prediction task as a generalised link prediction task in our evaluation. We focus on fundamental triangle prediction task, though it can be generalised to other motif structures. GCN and H$^2$NT-G are not studied here since GCN is primarily designed for  node classification tasks.

In our experiments, we first randomly remove some triangles to be used as testing set. The number of removed triangles are shown in Table \ref{statistics} (the right most column). Then we train all models in the rest 
of the network. Note that the summation of the number of
triangles in testing and training are not equal with the total number of triangles
in the original graph since
triangle structures are correlated with each other in a network. 
To evaluate the performance, 
we take the following five steps:
1) Positive sampling: we sample existing triangles (i.e., testing set) in the original graph.
2) Negative sampling: we sample three-node tuples and ensure every tuple cannot compose triangles in the original graph. Its quantity is ten times over positive samples.
3)  After obtaining embedding of nodes, we calculate the mean of tuplewise similarity (e.g., dot product)  in positive and negative sampling sets.
4) Mix and sort similarity of negative and positive together, and use precision@$N_p$
to evaluate. Here, we set the maximal $N_p$ as 500 for all traffic networks and 10,000  for all Facebook social networks (noting the total number of triangles is at exponential scale) and with the reasoning that for a good model, the similarity of positive sampling should be larger than that of negative sampling. 
5)  Calculate all precision@$N_p$  from 1 to maximum $N_p$ and average them. Finally, the average results of 5 runs are reported in Table \ref{tab:motif-pred}.

We have following observations: 
1) H$^2$NT-A 
achieves the overall best performance over all datasets, and AROPE achieves
the second best. 
2) H$^2$NT-A improves the AROPE by 2.23\% on average.
Additionally, H$^2$NT-D can improve DeepWalk by 23.9\% for all
social networks on average, and it even can improve more than ten times for two sparse USA and Europe traffic networks. It shows the effectiveness of H$^2$NT for preserving triangle structures in embedding space.


\textbf{Node Classification.}
We evaluate the node classification performance. 
Specifically, we randomly select a portion of nodes as training set 
 and leave the rest as test set. Then, we train a one-vs-all logistic regression with L2 regularisation. We repeat the process for 10 times and report the average accuracy in  Table \ref{tab:classification}. We have following observations:
1) H$^2$NT-D achieves the overall best performance. 
2) H$^2$NT-D and H$^2$NT-A have poorer results than DeepWalk and AROPE, i.e., there is degradation rather than improvement. The reason could be matrix factorisation and convolutional neural network are less sensitive to heterophily. 

\textbf{Structural Role Classification.}
Earlier, we heuristically show that the H$^2$NT can help preserve node centrality as an application of the structural role classification. To validate the effectiveness, we conduct this task on Brazil, USA and Europe and show result in Table \ref{tab:role-class} with 90\% training ratio.

We observe that: 1) H$^2$NT-A and H$^2$NT-D achieve the best performance on Europe
and USA respectively. Struc2vec is
specifically designed to this task so it achieves better performance on Brazil than
H$^2$NT-based 
methods. It could be caused by sparsity problem of H$^2$NT-based methods due to motif representation, and especially for Brazil, the most sparse network among all datasets. 2) H$^2$NT-based method can  improve the overall performance of original methods. 
For example, in Europe, H$^2$NT-D, H$^2$NT-A and H$^2$NT-G improve 17.8\%, 7.9\% and 24.3\% over original DeepWalk, AROPE, and GCN respectively.

\textbf{Computational Efficiency.}\label{sec:com-eff}
Table \ref{tab:time} compares the computational time of the original AROPE, DeepWalk and GCN with
H$^2$NT-A, H$^2$NT-D and H$^2$NT-G. For our H$^2$NT, the computational time includes the whole pipeline from input original network to output network embedding.    We have two key observations:
1) Our H$^2$NT-A, H$^2$NT-D and H$^2$NT-G can improve efficiency of AROPE, DeepWalk, and GCN by 35.1\%, 45.7\% and 10.1\%
respectively. This is because our H$^2$NT only use local motif structures and sparsify the original graph, which accelerates the optimisation process of combined methods and improves efficiency.
2) We further study the overhead of the motif representation calculation. Our studies show that the motif calculation is not the most important part in computational cost, e.g. it accounts for only 5\% and 8\% of the total time of H$^2$NT-A on Amherst and Rochester, respectively. Thus, the efficiency gain due to increased sparsity has exceeded this small overhead, leading to an overall improvement of computational efficiency.

\begin{figure} [!t] 
\centering
\subfigure{\includegraphics[width=40mm]{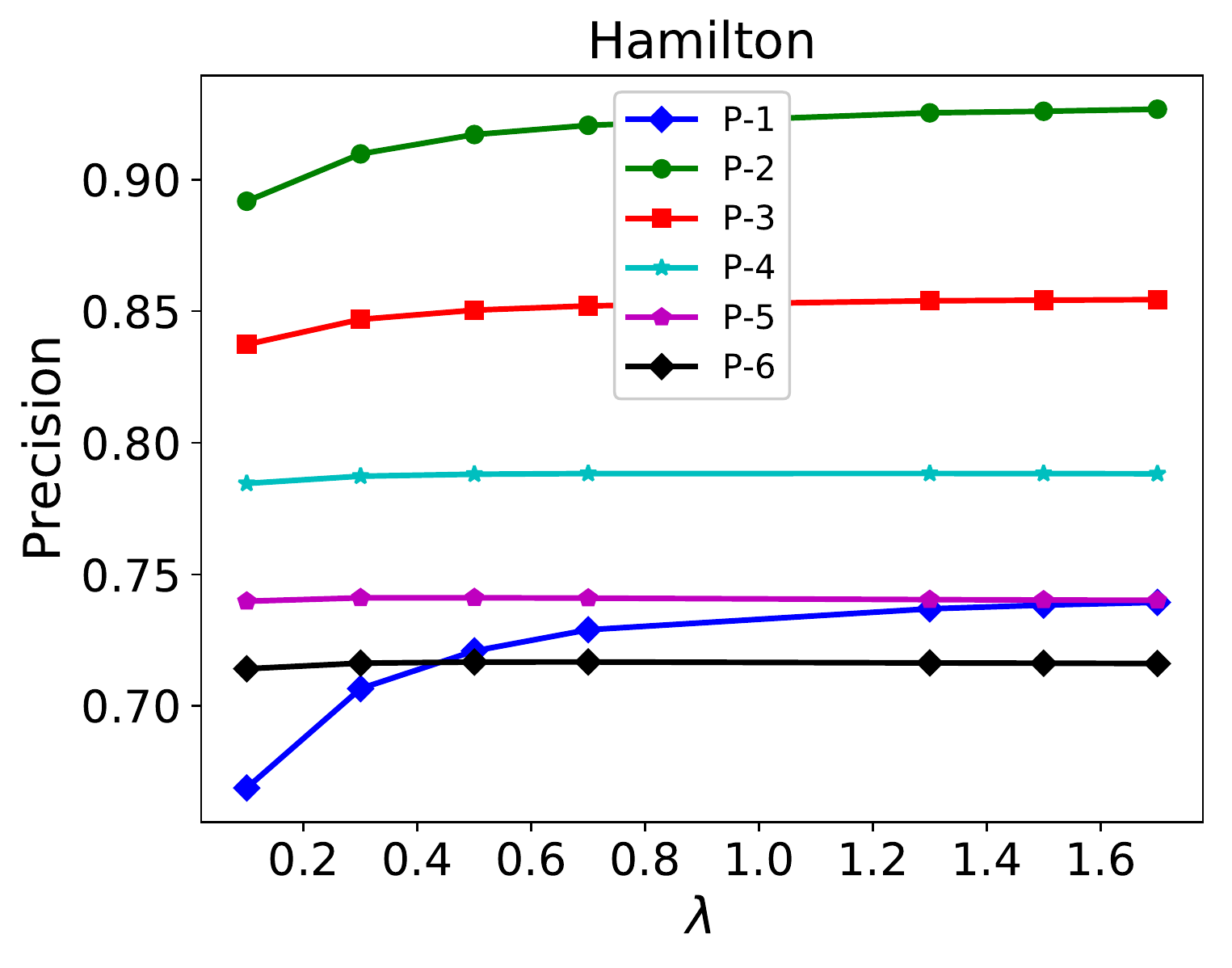}\hspace*{-0.5em}\label{sensitivity:subfig1}}
\subfigure{\includegraphics[width=40mm]{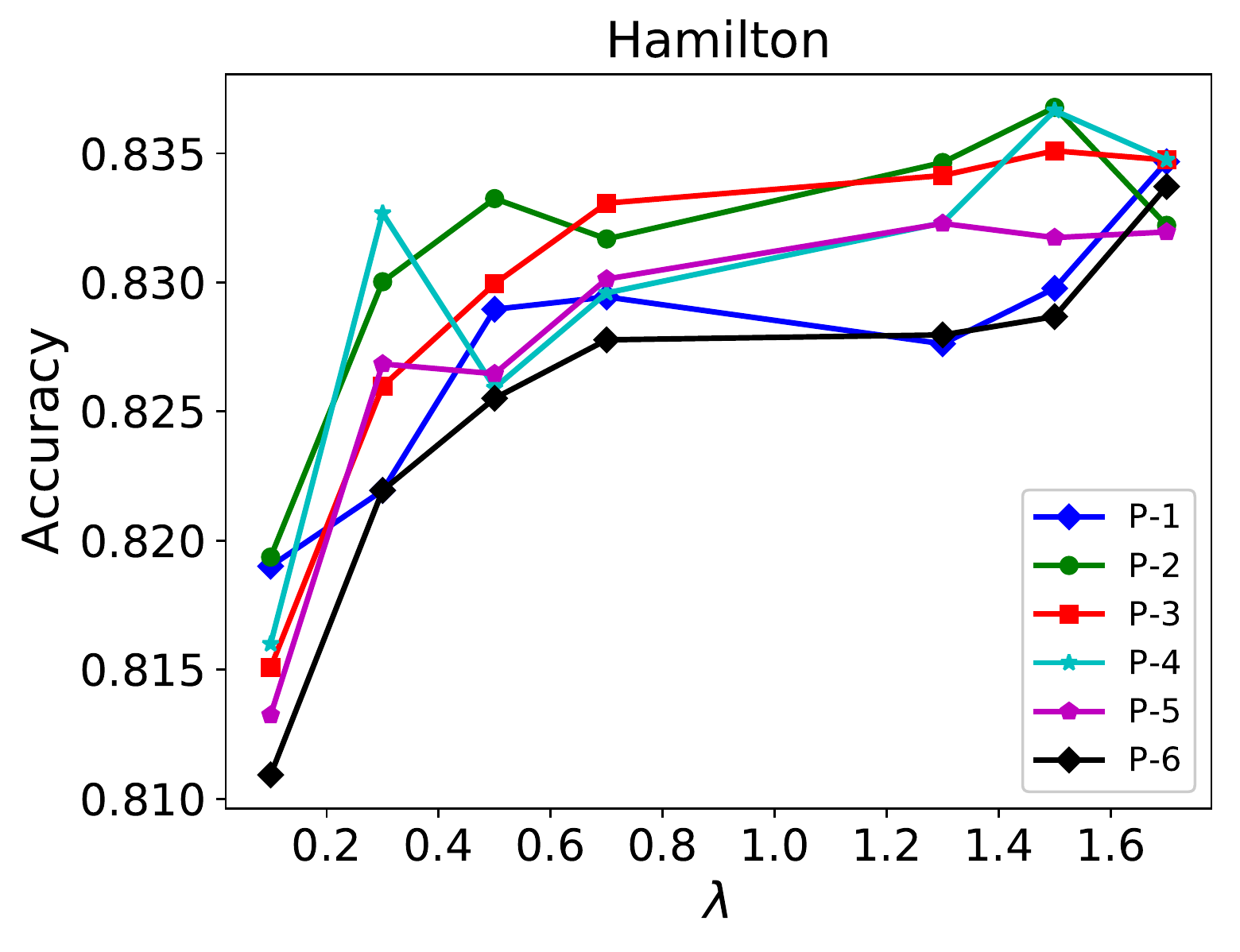}\hspace*{-0.7em}\label{sensitivity:subfig2}}
\vspace{-6.7mm}
\caption{Sensitivity analysis on two hyperparameters proximity order $P$ and
unifying weight $\lambda$. Left: H$^2$NT-A for motif prediction on Hamilton. Right: H$^2$NT-D for node classification on Hamilton. }
\label{fig:sensitivity}
\vspace{-1mm}
\end{figure}

\textbf{Sensitivity Analysis.}
We conduct a sensitivity study for two hyperparamters:  the order of HOP $P$ and unifying weight $\lambda$, as shown in Fig. \ref{fig:sensitivity}. The left one shows the  performance variation of H$^2$NT-A for motif
prediction task on Hamilton. We see that H$^2$NT-A is less sensitive to the unifying weight
than the number of HOP. The right one
shows the performance variation of H$^2$NT-D for node classification task on Hamilton. 
We see that  H$^2$NT-D is more sensitive to the unifying weight than the number
of HOP.

\begin{table}[t!]
	\caption {Computational time (in seconds). The last row shows the average with the most efficient result in \textbf{bold} and the second  \underline{underlined}. }\label{tab:time}
	\vspace{-3mm}
	\setlength{\tabcolsep}{1.4pt}
	\renewcommand{\arraystretch}{1}
 	\resizebox{\columnwidth}{!}{
	\centering
	\large
	\begin{tabular}{lcc||cc||ccccc}
		\toprule 
		Datasets &AROPE&H$^2$NT-A&DeepWalk& H$^2$NT-D&GCN&H$^2$NT-G\\ 
		\midrule	
		Amherst  &3.99 &2.79	&216.97	&99.17 &26.46			&28.34	\\
		Hamilton &4.06 &2.84	&218.87	&112.23&32.57			&28.10	\\
		Mich     &5.35 &2.63	&306.91	&121.21&25.46			&19.89	\\
		Rochester&5.40 &3.96	&443.54	&311.06&72.17			&64.42	\\ \hline
		Average &\underline{4.70} &\textbf{3.05}	&296.57	&160.92 &39.16			&35.19\\
		\bottomrule
	\end{tabular}
 	}
 	\vspace{-4mm}
\end{table}

\section{Conclusion}
In this paper, we proposed an H$^2$NT framework that 
 makes use  of  motif  representations  to  transform  a  network  into  a  new  network  preserving both homophily and heterophily  via flexible and complementary  micro-level and macro-level walk paths. 
 H$^2$NT can
    be integrated with any existing network embedding methods without requiring
    changing their cores such that it can take advantage of powerful network embedding methods proposed recently.
    We conducted experiments on node classification, 
    structural role classification and newly designed motif (link) prediction to show the
    superior prediction performance.

\noindent
\textbf{Acknowledgement.}
This work is supported by the Amazon Research Awards 2018.


\bibliographystyle{named}

\bibliography{ijcai20}
\newpage


\section*{Supplementary material}

# Supplementary Material

*Unifying Homophily and Heterophily Network Transformation via Motifs*

# Lemmas and Proofs

**Proofs about Relationship between Homophily and HOP**

**Lemma 1.** *Let $G_f \sim G_f(mk, k, p, q)$ be a fully-connected PPM with $k$ clusters $C = \{C_1, \cdots, C_i, \cdots, C_k\}$, nodes $\{v_i, v_j\} \in C_s$ and $v_k \in C_t$, the value of $l^{th}$-order proximity between $v_i$ and $v_j$ is $a_{ij}^{(l)}$, then,*

$$a_{ik}^{(l)} < a_{ij}^{(l)}, \qquad (1)$$

*Proof.* Based on Chapman-Kolmogorov equations (Tijms 2012), we have

$$a_{ik}^{(l)} = \sum_{z=1}^{kn} a_{iz}^{(l-1)} \cdot a_{zk}, \ a_{ij}^{(l)} = \sum_{z=1}^{kn} a_{iz}^{(l-1)} \cdot a_{zj} \qquad (2)$$

The above equations interpret the proximity that it is going from node $v_i$ to node $v_j$ in $l$ steps is obtained by summing the proximity of the mutually events of going from node $i$ to some node $v_k$ in the first $n-1$ walk steps and then going from node $k$ to node $j$ in the $l^{th}$ walk step. Therefore, the difference $a_{ij}^{(l)} - a_{ik}^{(l)} = \sum_{z \in C_s} a_{iz}^{(l-1)} \cdot (a_{iz} - a_{zk}) + \sum_{z \in C_t} a_{iz}^{(l-1)} \cdot (a_{iz} - a_{zk}) + \sum_{z \in \{C_h | 1 \leq h \leq k, h \notin \{s,t\}\}} a_{iz}^{(l-1)} \cdot (a_{iz} - a_{zk}) = \sum_{z \in C_s} a_{iz}^{(l-1)} \cdot (p - q) - \sum_{z \in C_t} a_{iz}^{(l-1)} \cdot (p - q)$.

Then, according to Lemma 2 and Lemma 3,

**Lemma 2.** *Let $G_f \sim G_f(nk, k, p, q), \{v_i, v_d, v_e\} \in C_s$, then we have $a_{id}^{(h)} = a_{ie}^{(h)}$.*

**Lemma 3.** *Let $G_f \sim G_f(nk, k, p, q), v_i \in C_s, \{v_r, v_s\} \in C_t$, we have $a_{ir}^{(h)} = a_{is}^{(h)}$.*

we can have $\sum_{z \in C_t} a_{iz}^{(l-1)} = n a_{ij}^{(l-1)}$ and $\sum_{z \in C_s} a_{iz}^{(l-1)} = n a_{ik}^{(l-1)}$.

Therefore,

$$a_{ij}^{(l)} - a_{ik}^{(l)} = (p-q)(n a_{ij}^{(l-1)} - n a_{ik}^{(l-1)})$$
$$= n^{l-1}(p-q)^{l-1}(a_{ij} - a_{ik}) = n^{l-1}(p-q)^l > 0.$$

This completes the proof of the inequality Eq. (1). □

**Proofs about Effectiveness of Homophily Representation**

**Lemma 4.** *Let $G \sim G(mk, k, p, q)$ be an unweigthted graph sampled by a generalized PPM, $\{v_i, v_j\} \in C_i$ and $\{v_k\} \in C_j$, $\{v_d\} \in C_b$, $j \neq b$ and $(v_i, v_j) \in E, (v_d, v_k) \in E$. We use $\mathbb{E}[w(v_i, v_j)]$ to indicate the expected number of triangles containing the edge $\{v_i, v_j\}$, then*

$$\mathbb{E}[w(v_d, v_k)] < \mathbb{E}[w(v_i, v_j)]. \qquad (3)$$

*Proof.* For the case of the edge $(v_i, v_j)$:

$$\mathbb{E}[w(v_i, v_j)] = (n-2)p^2 + (k-1)nq^2,$$

The first term $(n-2)p^2$ corresponds to triangles where the third vertex is in the same component as $v_i$ or $v_j$, the second term $(k-1)nq^2$ to triangle where the third vertex is in another component.

For the case of edge $(v_i, v_k)$:

$$\mathbb{E}[w(v_i, v_k)] = 2(n-1)pq + (k-2)nq^2$$

The first term $2(n-1)pq$ corresponds to triangles where the third vertex is in the same component as $v_i$ or $v_k$, the second term $(k-2)nq^2$ to triangle where the third vertex is in another component. For simplicity we avoid lower order terms as asymptotically the difference does not matter. Thus the difference between $\mathbb{E}[w(v_i, v_j)]$ and $\mathbb{E}[w(v_i, v_k)]$

$$\mathbb{E}[w(v_i, v_j)] - \mathbb{E}[w(v_i, v_k)] =$$
$$np^2 + (k-1)nq^2 - 2npq - (k-2)nq^2 = n(p-q)^2 > 0$$

This completes the proof of the inequality Eq.(3). □

# References

proximity preserved network embedding. In *KDD*, pages 2778–2786. ACM, 2018.

Tijms, H. 2012. *Understanding probability*. Cambridge University Press.


\end{document}